\documentclass[twocolumn,prl,aps]{revtex4}
\usepackage{amsmath}
\usepackage{amssymb}
\usepackage{graphicx}
\usepackage{dcolumn}

\begin{document}

\preprint{Phys.Rev.B}

\title{Vorticity induced negative nonlocal resistance in viscous two-dimensional electron system.}
\author{A. D. Levin,$^1$ G. M. Gusev,$^1$ E. V. Levinson,$^1$ Z.D. Kvon,$^{2,3}$
 and A. K. Bakarov $^{2,3}$}

\affiliation{$^1$Instituto de F\'{\i}sica da Universidade de S\~ao
Paulo, 135960-170, S\~ao Paulo, SP, Brazil}
\affiliation{$^2$Institute of Semiconductor Physics, Novosibirsk
630090, Russia}
\affiliation{$^3$Novosibirsk State University, Novosibirsk 630090,
Russia}

\date{\today}
\begin{abstract}
We report non-local electrical measurements in a mesoscopic size two-dimensional (2D)
 electron gas in a GaAs quantum well in a hydrodynamic regime. Viscous electric flow is expected to be dominant
  when electron-electron collisions occur more often than the impurity or phonon scattering events.
We observe a negative nonlocal resistance and attribute it to the formation
of whirlpools in the electron flow. We use the different nonlocal transport geometries and compare the results with a theory demonstrating the
significance of hydrodynamics in mesoscopic samples.


\end{abstract}

\maketitle

\section{Introduction}
It is generally believed that, in the absence of disorder, a many-body electron system resembles the viscous flow. Hydrodynamic characteristics can be specially enhanced in a pipe flow setup, where the mean free path for electron-electron
collision, $l_{ee}$, is much shorter than the sample width $W$, while the mean free path due to impurity and phonon scattering, $l$, is larger than $W$. Viscosity is characterized by momentum relaxation in the fluid and, in narrow samples, occurs
at the sample boundary. Calculation of the shear viscosity, $\eta$, is a difficult task because it requires knowledge of particle interactions on the scale of $l$ [1].

It has been predicted that the resistivity of metals in the hydrodynamic regime is proportional
to electron shear viscosity, $\eta=\frac{1}{4}v_{F}^{2}\tau_{ee}$, where $v_{F}$ is the Fermi velocity and $\tau_{ee}$ is the
electron-electron scattering time $\tau_{ee}=l_{ee}/v_{F}$ [2-6].
This dependency could lead to interesting properties. For example, resistance  decreases with the square of temperature,
$\rho \sim \eta \sim \tau_{ee} \sim T^{-2}$, the so called Gurzhi effect, and with the square of sample width
$\rho \sim W^{-2}$. The negative differential resistance has been observed previously in GaAs wires, which has been interpreted as the Gurzhi effect due to heating by the current [7].
\begin{figure}[ht]
\includegraphics[width=8cm]{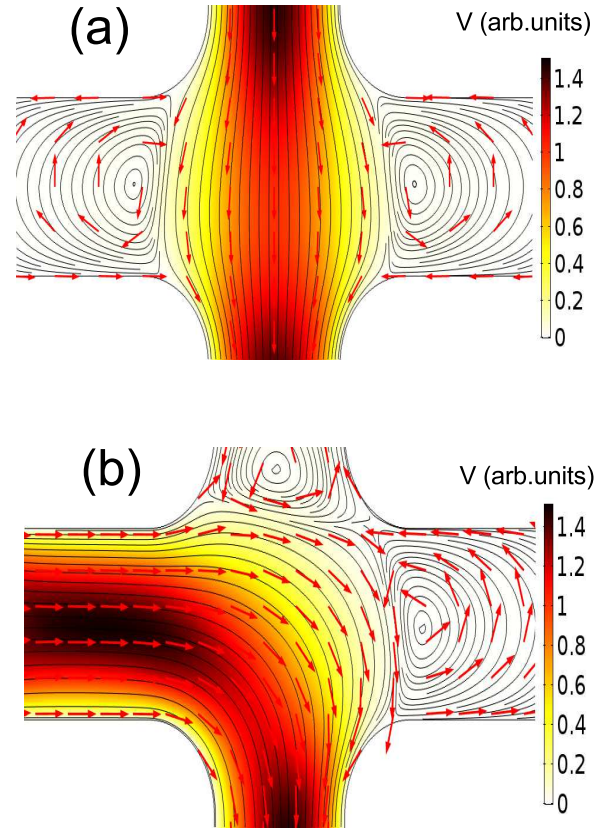}
\caption{(Color online) Sketch of the different transport set up measurements, showing a velocity flow profile.
(a) Non local transport set up, proposed in paper [8].
(b) Non local (vicinity) transport set up, proposed in papers [8-11].}
\end{figure}
A remarkable manifestation of the hydrodynamic effect is a swirling feature in the flow field, referred to as a vortex.
The vorticity can drive the current against an applied  electric field and generate backflow near the current injection region,
which can be detected in the experiment as a negative voltage drop [8]. A different transport measurement set up has been proposed for the identification
of viscosity related features in the hydrodynamic regime [8-11].

When fluid flows along a pipe, a quadratical velocity profile is formed, which leads to the Gurzhi effect, and can be detected
from the anomalous temperature and sample width dependence, as is mentioned above. For illustration  we modeled the Poiseuille flow for a two dimensional
neutral fluid. Fig. 1a shows the configuration, which has been proposed in paper [8],
and where the current is injected across the sample between vertical probes. In this geometry, one can see the vortex or whirpools in the liquid flow
outside of the main current path. As a consequence, for an electronic fluid, a negative voltage drop occurs  across the strip in close proximity to the current probes.
Fig.1b illustrates the nonlocal, vicinity transport geometry, where the current is injected in the left lateral and bottom contacts, while the voltage drop occurs near
the current injection region. This geometry has been proposed in papers [9-11], and the model clearly demonstrates the formation of whirpools in the hydrodynamic flow, yielding a negative nonlocal signal in transport measurements [11]. Note that the swirling features can be observed only in the nonlocal configuration.

The nonlocal vicinity effect has been studied experimentally in an ultracleen graphene sheet [11]. It has been demonstrated that the
nonlocal signal undergoes a sign change from positive, at low temperatures, to negative, above elevated temperatures, that is associated
with whirpool emergence in the hydrodynamic regime. Near room temperature, the signal again undergoes a sign change because the Ohmic contribution
starts to dominate the vicinity response at high T. Note that such dramatic experimental appearance of hydrodynamic features
in nonlocal transport has not been accompanied by observation of the Gurzhi effect in local transport. Moreover, the transversal nonlocal
geometry (Fig.1) has not been studied experimentally with respect to possible vorticity effects. Other materials, such as GaAs quantum wells, have
a particular interest because they possess the highest mobility over wide temperature ranges. It is also worthwhile to extend the theoretical
approach [8-11] to a two-dimensional electron in GaAs with a parabolic energy spectrum, which is different from the linear spectrum in Graphene.

A series of updated theoretical approaches has been published recently [12-15], providing additional possibilities to determine the viscosity from magnetotransport measurements, which can be used for comparison with non local measurements.

In this study, we measure the nonlocal resistance in mesoscopic GaAs quantum well systems. We determine all relevant electron viscous parameters from the longitudinal
magnetoresistance in a wide temperature range, which provides an estimate of the nonlocal signal, and compare it with experimental results.
A good qualitative agreement between experimental and simulated data has been obtained.

\begin{figure}[ht]
\includegraphics[width=9cm]{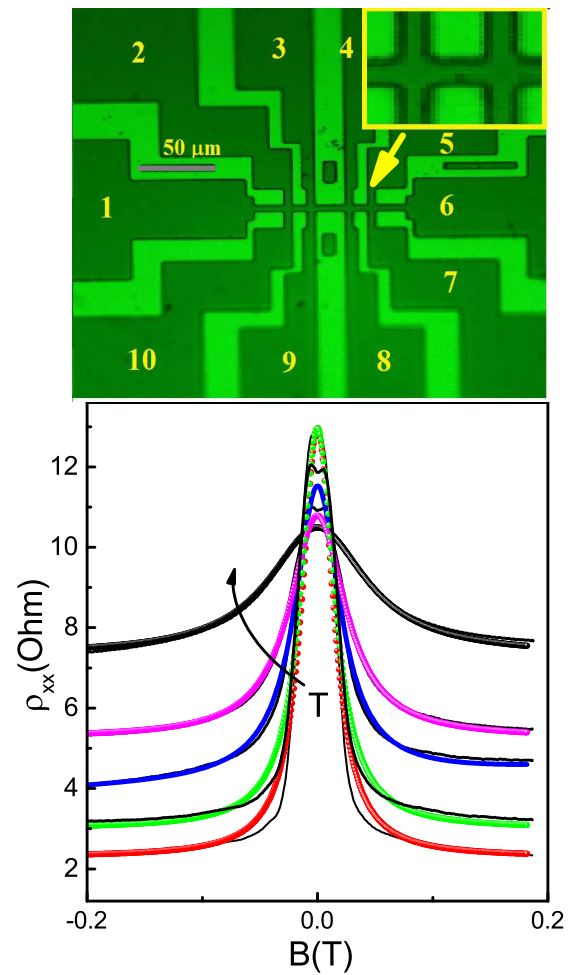}
\caption{(Color online)
Top -image of the Hall-bar device. Left top-zoomed  Hall bar bridge.
Temperature dependent magnetoresistance of a GaAs quantum well in a Hall bar sample, $W=5 \mu m$. Thick lines are examples illustrating magnetore-sistance calculated from Eqs. 1,2 for different temperatures: 4.2 K (red), 14 K(green), 19 K (blue), 26 K (magent) and 37.1 K (black). }
\end{figure}

\begin{figure}[ht]
\includegraphics[width=9cm]{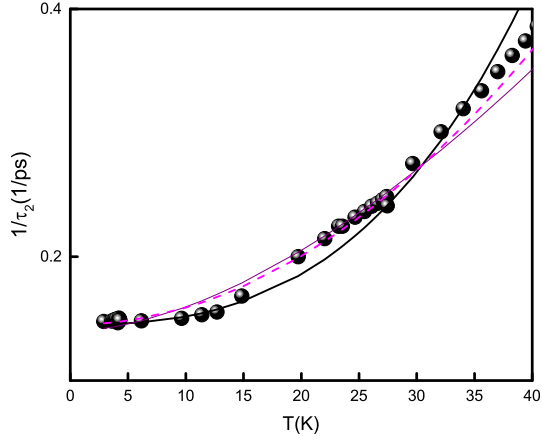}
\caption{(Color online)
Relaxation rate $1/\tau_{2}$ as a function of the temperature obtained by fitting
the theory with experimental results, $W=5 \mu m$. Thick black line is Equation 2, black line is Equation 3, dashes are Equation 4.}
\end{figure}

\section{Negative giant magnetoresistance, experiment and discussion}
 Our samples are high-quality, GaAs quantum wells with a width of 14~nm, 
with electron density $n_{s}=6\times10^{11} cm ^{-2}$ and a mobility of $\mu=2.5\times10^{6} cm^{2}/Vs$ at T=1.4K.
Other parameters characterizing the electron system are given in Table 1.
We present experimental results on Hall-bar devices designed
in two different configurations.
Design I  consists of three, $5 \mu m$ wide consecutive segments of
different length ($10, 20 , 10 \mu m$), and 8 voltage probes. Fig.2 (top) shows the image of a typical multiprobe Hall device I.
Design II is also a Hall bar with three, $2 \mu m$ wide consecutive segments of
different length ($2, 7 , 2 \mu m$), and 8 voltage probes.
The measurements were carried out in a
VTI cryostat, using a conventional
lock-in technique to measure the longitudinal $\rho_{xx}$ resistivity with an
ac current of $0.1 - 1 \mu A$ through the sample, which is
sufficiently low to avoid overheating effects.
5 Hall bars from two different wafers were studied.

\begin{table}[ht]
\caption{\label{tab1} Parameters of the electron system in a mesoscopic samples at $T=1.4 K$. Parameters are defined in the text.}
\begin{ruledtabular}
\begin{tabular}{lcccccc}
&W&$n_{s}$ & $v_{F}$ & $l$  & $l_{2}$ & $\eta$\\
&($\mu m$) & $(10^{11} cm^{2}$) & $(10^{7} cm/s)$ &  $(\mu m$) & $(\mu m$) & $(m^{2}/s)$\\
\hline

&$5$& $9.1$  & $4.1$ &  $40$ & $2.8$ & $0.3$\\
&$2$& $6.0$  & $3.3$ &  $20.6$ & $1.4$ & $0.12$\\
\end{tabular}
\end{ruledtabular}
\end{table}
Longitudinal magnetoresistance has been studied in previous research for different configurations of the current and voltage probes [16].
Before analyzing the nonlocal effect and in order to make this analysis more complete, we present the results of measurements of the longitudinal magnetoresistivity $\rho_{xx}(B)$.
Fig. 2a shows $\rho_{xx}(B)$ as a function of magnetic field and temperature.
One can see two characteristic features: a giant negative magnetoresistance $(\sim 400-600 \%)$ with a Lorentzian-like shape, and a pronounced temperature dependence on zero field resistance.
In the hydrodynamic approach, the semiclassical treatment of the transport describes the motion of carriers when the higher order moments of the distribution function are taken into account. The momentum relaxation rate $1/\tau$ is determined by electron interaction with phonons and static defects (boundary).
The second moment relaxation rate $1/\tau_{2,ee}$ leads to the viscosity and contains the contribution
from the electron-electron scattering and temperature independent scattering by disorder [12,13].
It has been shown that conductivity obeys the additive relation and is determined by two independent $\textit{parallel}$ channels:
the first is due to momentum relaxation time and the second due to viscosity [12,13].
This approach allows the introduction of the magnetic field dependent viscosity tensor and the derivation of the magnetoresisivity tensor [12-15]:
\begin{equation}
\rho_{xx}= \rho_{0}^{bulk}\left(1+\frac{\tau}{\tau^{*}}\frac{1}{1+(2\omega_{c}\tau_{2,ee})^{2}}\right),\,\,\,
\end{equation}

where $\rho_{0}^{bulk}=m/ne^2\tau$, $\tau^{*}=\frac{W(W+6l_{s})}{12\eta}$, viscosity $\eta=\frac{1}{4}v_{F}^{2}\tau_{2,ee}$.

We also collect the equations for relaxation rates separately:
\begin{equation}
\frac{1}{\tau_{2,ee}(T)}=A_{ee}^{FL}\frac{T^{2}}{[ln(E_{F}/T)]^{2}}+\frac{1}{\tau_{2,0}},
\end{equation}

where $E_{F}$ is the Fermi energy, and the coefficient $A_{ee}^{FL}$  be can expressed via the Landau interaction parameter, however, it is difficult to calculate quantitatively (see discussion in [12]).
The relaxation rate $\frac{1}{\tau_{2,0}}$ is not related to the electron-electron collisions, since any process responsible for relaxation of the second moment of the distribution function, even scattering by static defect, gives rise to viscosity [12].
A logarithmic factor is also present in the expression for quantum lifetime of weakly interacting 2D gas due to electron-electron scattering [17]:
\begin{equation}
\frac{\hbar}{\tau_{0,ee}(T)}=A_{ee}^{0}\frac{T^{2}[ln(2E_{F}/T)]}{E_{F}}+\frac{\hbar}{\tau_{2,0}},
\end{equation}

where $A_{ee}^{0}$ is a numerical constant of the order of unity. Note, however, that since the relaxation time $\tau_{0,ee}$ is related to the kinematic of the electron-electron collisions,
Expression 2 is more convenient and it is preferable to use.  Finally, it has been shown that due to the disorder assisted contribution to the relaxation rate of
the second moment of the distribution function, the expression is rewritten as:
\begin{equation}
\frac{1}{\tau_{2,ee}^{da}(T)}=A_{ee}^{da}T^{2}+\frac{1}{\tau_{2,0}},
\end{equation}
where the coefficient $A_{ee}^{da}$  depends on the disorder type and its strength [12].
The moment relaxation rate is expressed as:
\begin{equation}
\frac{1}{\tau}=A_{ph}T+\frac{1}{\tau_{0}},
\end{equation}
where $A_{ph}$ is the term responsible for the phonon scattering [18,19], and $\frac{1}{\tau_{0}}$ is the scattering rate due to static disorder
(not related to the second moment relaxation rate $\frac{1}{\tau_{2,0}}$).

We fit the magnetoresistance curves in Fig. 2 and the resistance in zero magnetic field  with the 3 fitting parameters :
$\tau(T)$, $\tau^{*}(T)$ and $\tau_{2,ee}(T)$.
Fig.3 shows the dependencies of $1/\tau_{2,ee}(T)$  extracted from the comparison of the magnetoresistance shown in Figure 2 and Equation 1.
We compare the temperature dependence of  $\frac{1}{\tau_{2,ee}(T)}$  with theoretical predictions given by Equations 2-4 and present the results of such comparison in Figure 3.
The following parameters are extracted:
$1/\tau_{2,0}=1.45\times10^{11}$ s, $A_{ee}^{FL}=0.9\times10^{9} s^{-1}K^{-2}$, $A_{ee}^{0}=1.3$,  $A_{ee}^{da}=2.0\times10^{10} s^{-1}K^{-2}$.
All theoretical curves demonstrate reasonable agreement  within experimental uncertainty.
Hence, these mechanisms lead to nearly equivalent results and cannot be unambiguously distinguished based only on the temperature dependence of the relaxation time.
Note, that analysis of the nonlocal effect, considered below, does not depend on the relaxation mechanism.

In addition we extract the temperature dependence of the moment scattering rate and determine parameters  $A_{ph}=10^{9}sK^{-1}$ and $\tau_{0}=5\times10^-{10} s$,
which are correlated with previous studies [17,18]. Relaxation time  $\tau^{*}(T)$ depends on the  $\tau_{2,ee}(T)$ and boundary slip length $l_{s}$.
Comparing these values, we find that $l_{s}=3.2 \mu m < L$, and, therefore, in our case, it is appropriate to use diffusive boundary conditions.
Table 1 shows the mean free paths : $l=v_{F}\tau$, $l_{2}=v_{F}\tau_{2,ee}$  and viscosity, calculated with parameters extracted from the fit of experimental data.

\section{Experiment: nonlocal resistance}
In this section, we focus on the nonlocal configurations because such geometry facilitates the observation of current whirpools.
Fig. 4 shows the transport in a nonlocal set up, where the current is injected across the strip between probes 4 and 8.
The voltage drop is measured between probes 5 and 7. Below we refer to it as C1 configuration. Poiseuille flow for a two dimensional liquid is presented in Figure 1a. Note, however, that 2D charged liquid  displays pronounced ballistic transport behaviour. One can see
strong oscillations in weak magnetic fields due to geometrical resonance effects considered in the semiclassical billiard model [20,21].
We perform numerical simulations of the electron trajectories in ballistic structures. The results of theses simulations (black dots) are compared to the experimental data. We observe an agreement with experimental data only at low magnetic field.
Although the position of the resistance peaks at higher magnetic field coincide with calculations, the negative peak has a much smaller value,
and the positive peak is wider than that obtained from the billiard model. Figure 4 also shows the evolution of the nonlocal magnetoresistance with temperature. One can see that all oscillations are smeared out by temperature and magnetoresistance at high temperature has a parabolic shape.
Remarkably, the nonlocal resistance at B=0 is positive at low temperatures, in accordance with the billiard model calculations, and then it changes sign and becomes negative at higher temperatures (Fig.6).
\begin{figure}[ht]
\includegraphics[width=9cm]{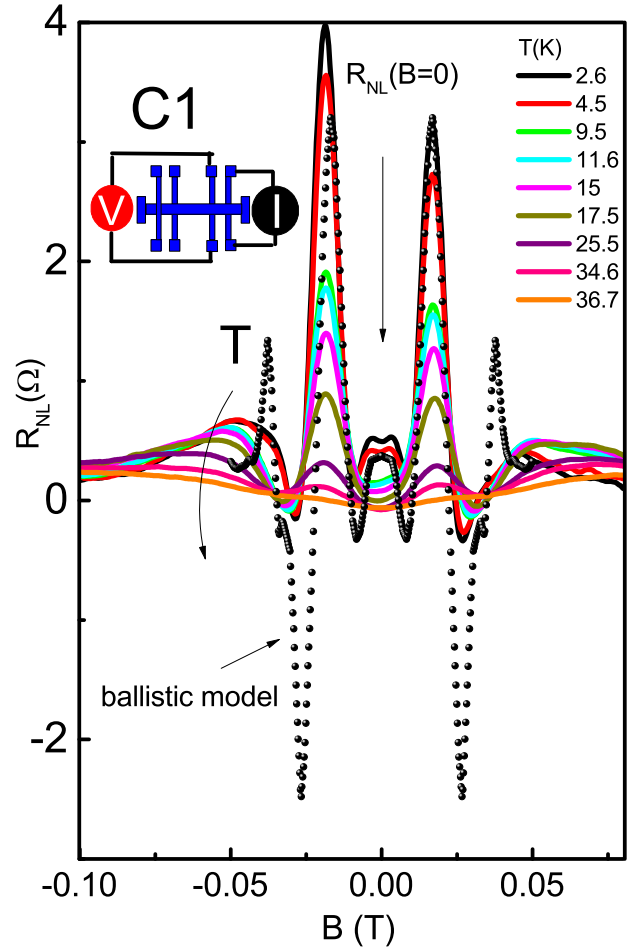}
\caption{(Color online)
Nonlocal transport signal versus magnetic field for different temperatures, $W=5 \mu m$.
 The dots represent results for the billiard model.
(b) T-dependence of the nonlocal signal for different sample configuration. Solid lines show the calculations from equation (2).}
\end{figure}
Fig. 5 shows the transport in a nonlocal set up, where the current is injected  between probes 1 and 8 and the voltage is measured between probes 5 and 6 (referred to as configuration C2).
The Poiseuille flow for a two dimensional liquid is presented in Figure 1 (b). As in configuration C1, one can see strong oscillations due to the geometrical resonance effect. Note that the ballistic transport in this configuration is very well established and studied previously in numerous publications [20,21]. In cross junction geometry, it was denominated as bend resistance [21]. We also perform the classical simulations for the transport
in configuration C2, and the results are displayed in Figure 5. Note, however, that in contrast to configuration C1, the bend resistance reveals a strong negative resistance peak near zero magnetic field [21,22]. This peak may mask the negative nonlocal signal due to viscosity,
and detailed comparison is required to examine the significance of the hydrodynamic effect at low and high temperatures.
Figure 7 presents the results of the nonlocal resistance temperature measurements in   configuration C2 in zero magnetic field.
One can see that the signal dramatically drops to zero in the $W=5 \mu m$ sample, and  resistance changes sign at high temperature in the $W=2 \mu m$ sample.
We also used a similar voltage measurement set up, where the current is injected  between probes 1 and 8 and the voltage is measured between probes 4 and 5 (referred to as configuration C3). The nonlocal resistance in configuration C3 at zero magnetic field is shown in Fig.7 for both samples designs.
\begin{figure}[ht]
\includegraphics[width=9cm]{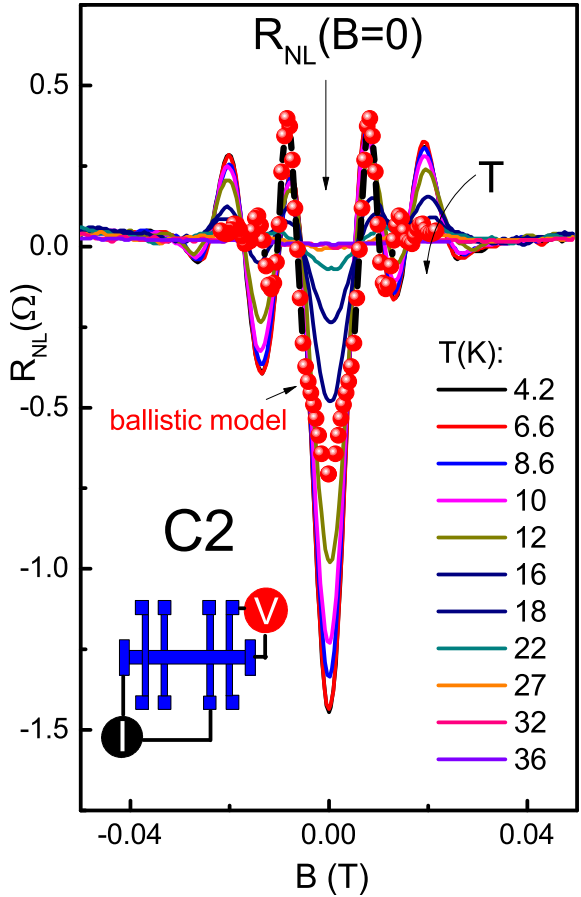}
\caption{(Color online)
Nonlocal transport signal versus magnetic field for different temperatures, $W=5 \mu m$.
The dots represent results for the billiard model.}
\end{figure}

\begin{figure}[ht]
\includegraphics[width=9cm]{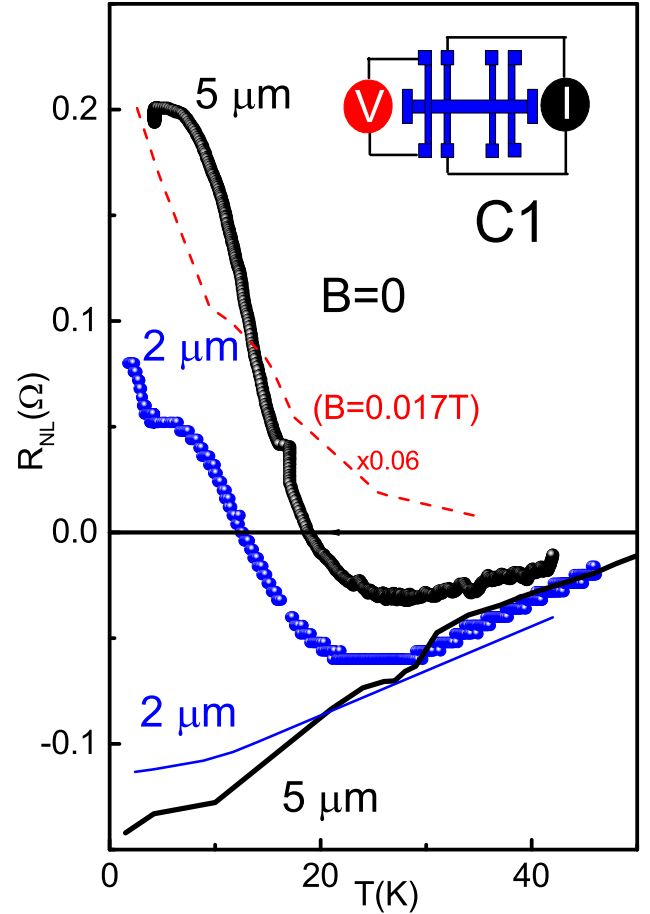}
\caption{(Color online)
T-dependence of the nonlocal signal for different sample configuration. Solid lines show the calculations from equation (6) for $x=10\mu m$ ($W=5 \mu m$) and $x=5 \mu m$ ($W=2 \mu m$).
Dashes-T dependence of the ballistic peak at $B=0.017 T$}
\end{figure}

\begin{figure}[ht]
\includegraphics[width=9cm]{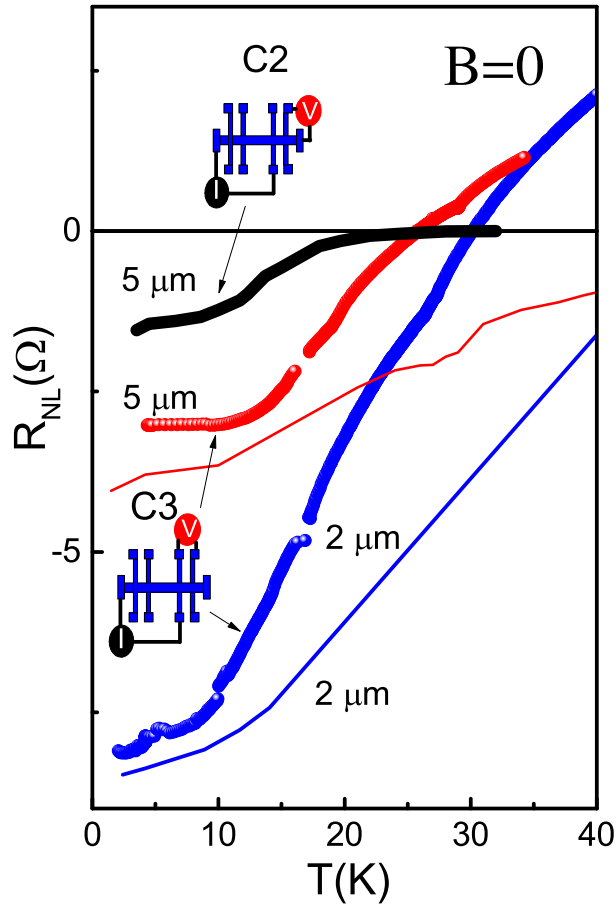}
\caption{(Color online)
T-dependence of the nonlocal signal for different sample configurations. Thin solid lines show the calculations from equation (7) for $x=3 \mu m$ ($W=5 \mu m$) and $x=1.5 \mu m$ ($W=2 \mu m$). }
\end{figure}

\section{Theory and discussion}
As has been shown in the previous chapter the viscosity leads to the incorporation of an extra relaxation mechanism [12-15] in zero magnetic field:
$\rho=\rho_{0}^{bulk}\left(1+\frac{\tau}{\tau^{*}}\right)$,
The dominant viscous contribution to resistivity corresponds to the small ratio between relaxation of the second moment of the electron distribution function and the first moment $\tau^{*}/\tau <<1$.

Comparative analysis between nonlocal geometries C1 and C2 demonstrates a qualitative difference. Crucially, the experimental observation of swirling features depends on the parameters that affect the spacial distribution of the two-dimensional potential inside the viscous charge flow. The first parameter is the boundary slip length $l_{s}$.
The boundary no-slip conditions correspond to the ideal hydrodynamic case of diffusive boundaries with $l_{s}=0$.
It has been shown that the negative nonlocal signal is robust to boundary conditions [10]. For example, the Gurzhi effect disappears  for free surface boundary conditions ($l_{s}=\infty$), while whirlpools in hydrodynamic electron flow, and the resulting negative nonlocal response, do exist.
The second parameter which drastically affects whirpool behaviour is the vorticity diffusion length $D_{\eta}=\sqrt{\eta\tau}$.
Fig. 8 represents the temperature dependence of characteristic lengths in a $W=5 \mu m$ sample. Previous studies have not investigated whether typically
developed current whirpools show sensitivity to the geometry and confinement effect [8-10]. However, the careful inspection of theoretical results [9]
reveals that geometry C1 exhibits the occurrence of whirpools only above the threshold value of
$D_{\eta}=0.225W$ (Figure 8). The vicinity geometry, C2, which is shown in Figure 1b, by contrary, allows the formation of current whirpools
for arbitrary small values of  $D_{\eta}$, but only in very close proximity to the current injector probe [10].  However, the value of $D_{\eta}$ affects the spatial extension of the whirpools, therefore, a high viscosity system facilitates observation of the negative
vicinity resistance for a voltage detector placed at a large distance from the current injection probe. Moreover, the ballistic effect may induce the negative vicinity signal [19] and, therefore, requires more careful qualitative analysis. In the previous section, we show the temperature dependence of low field magnetoresistance as well as the electrical resistivity
 over a temperature range extending from $1.7$ to $40 K$ and obtain variation of the viscosity time with temperature. We use this data to estimate the nonlocal signal in our samples. The models [8-10] predict negative nonlocal resistance in configuration C1 at the distance $\overline{x}=\pi x/W$ from the main current path
 in the limits of free surface boundary conditions ($l_{s}=\infty$)in zero magnetic field :
\begin{equation}
 R_{NL}^{C1}=-\rho_{0}\left\{\frac{ln[\tanh^{2}(\overline{x}/2)]}{\pi}+4\pi \left(\frac{D_{\nu}}{W}\right)^{2}\frac{\cosh(\overline{x})}{\sinh^{2}(\overline{x})}\right\},
\end{equation}

In contrast to configuration C1, the results for vicinity geometry can be simplified only in the limit where the distance between the current injection probe is infinite:
\begin{equation}
 R_{NL}^{C2}=-\frac{\rho_{0}}{2} \left\{\frac{ln[4\cal T]}{\pi}-\frac{\overline{x}}{W}+
 \pi \left(\frac{D_{\nu}}{W} \right)^{2}\frac{1}{\cal T} \right\},
\end{equation}
where ${\cal T}=\sinh^{2}(\overline{x}/2)$. Figure 9 shows the  nonlocal resistancies in both configurations as a function of distance between voltage probe and current injector $\overline{x}$ calculated from Equations 6 and 7 with parameters independently extracted from the local magnetoresistance measurements at $T=4.2 K$. For visualization of the data in the negative range, we used an absolute log scale.
 We observe that the magnitudes of nonlocal signals exhibit a universally exponential decay with distance from the current injector.
 Note that the nonlocal resistance is much stronger for geometry C1. The advantage of configuration C1 is that the ballistic contribution is positive and, therefore, it can be unambiguously discriminated from the negative viscous contribution.
 The calculated temperature dependence of $R_{NL}^{C1}$ is shown in Figure 6 for $x=10 \mu m$ ($W=5 \mu m$) and $x=5 \mu m$ ($W=2 \mu m$), which roughly correspond to the distance between the center of the probes.
Note that the ballistic contribution to the transport also depends on the temperature due to the thermal broadening of the Fermi distribution function and scattering by the phonons.
A rough estimate of the nonlocal ballistic resistance temperature dependence for $L<l$  may be obtained using the formula $R_{NL}\sim exp(-L/l)$, where $L$ is the distance between probes [23]. Fig.6 shows the T-dependence of the ballistic peak at $B=0.017 T$.
One can see a rapid decrease of the peak with temperature. Therefore, the negative nonlocal resistance in zero field and at high temperature can be attributed only to hydrodynamic effects.

\begin{figure}[ht]
\includegraphics[width=9cm]{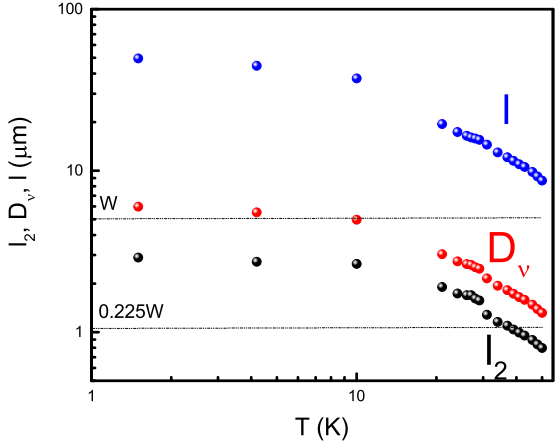}
\caption{(Color online)
The characteristic parameters as a function of the temperature for the sample with width $W=5 \mu m$. The whirpool threshold is indicated by the dashes.
}
\end{figure}
\begin{figure}[ht]
\includegraphics[width=9cm]{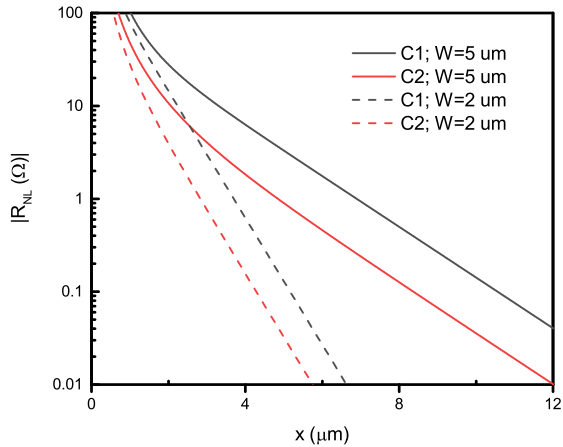}
\caption{(Color online)
The absolute value of the nonlocal resistance for two configurations as a function of the distance from the injector electrode, $T=4.2 K$, parameters are determined from
local magnetoresistance measurements.}
\end{figure}
We also compare predictions for configurations C2 and C3 with experimental results.
Note that we normalized ballistic resistance for the peak value at $B=0.008 T$ (Fig. 5), which we found more reliable, since this peak weakly depends on the boundary conditions and sample geometry [20]. The residual contribution at zero magnetic field could be due to viscous effects. In general, the ballistic contribution alone can explain the temperature dependence in zero field, below $20 K$, without taking into account the viscous term.
Above $T=20 K$, ballistic contribution should be exponentially small (see Fig. 6).
Figure 7 shows the calculations from Equation 7. Note that the analytical formula has been derived under several assumptions and we can apply the formula just for the evaluation of the upper limits of the signal. Figure 7 presents the results of such calculations.
One can see that the predicted signal agrees with experimental data for $x=3 \mu m$ ($W=5 \mu m$) and $x=1.5 \mu m$ ($W=2 \mu m$), which roughly correspond to the distance between the centers of the probes.
 Note that, in a realistic sample, the width of the probes is comparable with the sample width $W$, while the theory considers  $x << W$, also indicating the approximate character of the calculation. We may conclude here that geometry C1 exhibits a direct relation between the negative signal and formation of the current whirpools. In geometries C2 and C3,
negative nonlocal resistance follows the hydrodynamic predictions up to $30 K$, however, it is very likely that the ballistic contribution is comparable or bigger than the hydrodynamic one at low temperatures. Above $30 K$, we observe a positive signal, which disagrees with
both ballistic and hydrodynamic predictions. We attribute this behaviour to approaching the condition $D_{\eta}=0.225W$.
 Note that the observation of negative vicinity nonlocal resistance in graphene [11] requires more careful inspection of the ballistic contribution. Moreover, the condition $D_{\eta}=0.225W$ is not fully completed (see also discussion in [10]), therefore, our observation of the negative nonlocal resistance in geometry C1 provides more clear evidence of current vorticies. It is important to note that the transport signatures of the viscocity in the nolocal effect
are correlated in our samples with other observations, such as a giant longitudinal magnetoresistance and the Gurzhi effect [16].
\section{Summary and conclusions}
In conclusion, we have studied nonlocal transport in a mesoscopic two-dimensional electron system in terms of viscosity of the fluids.
In contrast to the Ohmic flow of the particles, viscous flow can result in a back flow of the current and negative nonlocal voltage.
We have measured voltage in different arrangements of current and voltage contacts and found a negative response, which we attributed to the formation of current whirlpools. Nonlocal viscosity-induced transport is strongly correlated with observations of the Gurzhi effect and low magnetic field transport described by hydrodynamic theory.

The financial support of this work by the Russian Science
Foundation (Grant No.16-12-10041), FAPESP (Brazil),
and CNPq (Brazil) is acknowledged.

\end{document}